\documentclass[english,preprint]{aastex}
\usepackage{lmodern}
\usepackage[T1]{fontenc}
\usepackage[latin9]{inputenc}
\usepackage{color}
\usepackage{amsbsy}
\usepackage{amssymb}
\usepackage{graphicx}
\usepackage{esint}
\usepackage{babel}
\begin{document}

\title{Turbulent Magnetohydrodynamic Reconnection Mediated by the Plasmoid
Instability}

\author{Yi-Min Huang\altaffilmark{1,2,3} }

\affil{Department of Astrophysical Sciences and Princeton Plasma Physics
Laboratory, Princeton University, Princeton, New Jersey 08543, USA}

\email{yiminh@princeton.edu}

\author{A. Bhattacharjee\altaffilmark{1,2,3} }

\affil{Department of Astrophysical Sciences and Princeton Plasma Physics
Laboratory, Princeton University, Princeton, New Jersey 08543, USA}

\altaffiltext{1}{ Princeton Center for Heliophysics}

\altaffiltext{2}{ Max Planck/Princeton Center for Plasma Physics}

\altaffiltext{3}{ Center for Magnetic Self-Organization in Laboratory and Astrophysical
Plasmas}
\begin{abstract}
It has been established that the Sweet-Parker current layer in high
Lundquist number reconnection is unstable to the super-Alfv\'enic
plasmoid instability. Past two-dimensional magnetohydrodynamic simulations
have demonstrated that the plasmoid instability leads to a new regime
where the Sweet-Parker current layer changes into a chain of plasmoids
connected by secondary current sheets, and the averaged reconnection
rate becomes nearly independent of the Lundquist number. In this work,
three-dimensional simulation with a guide field shows that the additional
degree of freedom allows plasmoid instabilities to grow at oblique
angles, which interact and lead to self-generated turbulent reconnection.
The averaged reconnection rate in the self-generated turbulent state
is of the order of a hundredth of the characteristic Alfv\'en speed,
which is similar to the two-dimensional result but is an order of
magnitude lower than the fastest reconnection rate reported in recent
studies of externally driven three-dimensional turbulent reconnection.
Kinematic and magnetic energy fluctuations both form elongated eddies
along the direction of local magnetic field, which is a signature
of anisotropic magnetohydrodynamic turbulence. Both energy fluctuations
satisfy power-law spectra in the inertial range, where the magnetic
energy spectral index is in the range from $-2.3$ to $-2.1$, while
the kinetic energy spectral index is slightly steeper, in the range
from $-2.5$ to $-2.3$. The anisotropy of turbulence eddies is found
to be nearly scale-independent, in contrast with the prediction of
the Goldreich-Sridhar theory for anisotropic turbulence in a homogeneous
plasma permeated by a uniform magnetic field.
\end{abstract}

\keywords{magnetic fields -- plasmas -- magnetohydrodynamics(MHD) -- magnetic
reconnection -- turbulence -- Sun:corona}

\section{Introduction}

Magnetic reconnection is a process that changes the topology of magnetic
field lines and releases magnetic energy in the form of plasma kinetic,
thermal, or nonthermal energy. It is widely believed to be the underlying
mechanism that powers explosive events such as geomagnetic substorms,
solar flares, coronal mass ejections (CMEs), gamma-ray bursts, as
well as sawtooth crashes in fusion plasmas \citep{Biskamp2000,PriestF2000,ZweibelY2009,YamadaKJ2010}.
Large-scale space and astrophysical environments where magnetic reconnection
takes place, such as solar corona, solar wind, interstellar medium,
molecular clouds, and accretion disks, are known to be turbulent \citep{Rickett1990,Narayan1992,BalbusH1998,LazarianVKYBD2012}.
Therefore, how turbulence and reconnection influence each other is
a question of great importance. Broadly speaking, the interplay between
turbulence and reconnection can be viewed from two complementary perspectives.
On the one hand, there can be many small-scale reconnection events
simultaneously taking place in a large-scale turbulent bath \citep{RappazzoVED2008,ServidioMSCD2009,ServidioMSDCW2010,RappazzoVE2010,ServidioDGWDCSCM2011,WanRMSO2014},
which may provide an energy source to heat the plasma, e.g. as in
the nanoflare scenario of coronal heating \citep{Parker1988}. On
the other hand, small-scale turbulence may also affect the reconnection
of a large-scale, coherent magnetic field \citep{LazarianV1999,KowalLVO2009,LoureiroUSCY2009,EyinkLV2011,LazarianEV2012,LazarianEVK2015a,LazarianEVK2015}.
This latter aspect may have significant implications in the energy
release of large-scale eruptions, such as coronal mass ejections (CMEs),
and is the main concern of this study. 

Traditionally, theoretical studies of magnetic reconnection have been
mostly focused on two-dimensional (2D) models (e.g. the classical
Sweet-Parker \citep{Sweet1958a,Parker1957} and Petschek \citep{Petschek1964}
models) under the assumption that reconnection occurs in a single,
stable current sheet. In recent years, there is growing evidence that
large-scale reconnection is likely to take place in a fragmented reconnection
layer, due to the presence of secondary instabilities, which are found
in a wide range of plasma models, including resistive magnetohydrodynamics
(MHD) \citep{Biskamp1986,ShibataT2001,LoureiroSC2007,Lapenta2008,BhattacharjeeHYR2009,CassakSD2009,HuangB2010,BartaBKS2011,HuangBS2011,ShenLM2011,LoureiroSSU2012,NiZHLM2012,HuangB2012,HuangB2013,Takamoto2013,WyperP2014},
Hall MHD \citep{ShepherdC2010,HuangBS2011}, and fully kinetic particle-in-cell
(PIC) simulations \citep{DaughtonSK2006,DaughtonRAKYB2009,DaughtonRKYABB2011,FermoDS2012,DaughtonNKRL2014}.
Among these secondary instabilities, the plasmoid (or secondary tearing)
instability in resistive MHD \citep{LoureiroSC2007,BhattacharjeeHYR2009}
has been extensively studied in recent years. Through high-resolution
2D simulations, it is now established that the plasmoid instability
leads to fast reconnection in resistive MHD with the reconnection
rate nearly independent of the resistivity \citep{BhattacharjeeHYR2009,HuangB2010,LoureiroSSU2012}.
When three-dimensional (3D) perturbations are allowed in a reconnection
configuration with a guide field, oblique tearing modes with resonant
surfaces (i.e. where $\mathbf{k}\cdot\mathbf{B}=0$) away from the
mid-plane can be excited in addition to the usual 2D modes \citep{BaalrudBH2012}.
In this work, we examine the possibility of establishing self-sustained
turbulent reconnection via the interaction between oblique tearing
modes, in contrast to previous MHD studies where turbulence is driven
through external forcing \citep{KowalLVO2009,LoureiroUSCY2009}. Realizing
self-sustained turbulent reconnection is essential for further comparison
with observations, as results from externally forced turbulent reconnection
will inevitably depend on the input power. Although self-sustained
turbulent reconnection through interacting oblique modes has been
reported in a recent fully kinetic collisionless PIC simulation by
\citet{DaughtonRKYABB2011}, the simulation system size is limited
to the order of several tens of ion skin depths, which is substantially
too small from an astrophysical point of view. To give an example,
typical values of ion skin depth in the solar corona are in the range
of $1$ -- $100$ meters, whereas the extent of a post-CME current
sheet can be more than $10^{9}$ meters \citep{CiaravellaR2008}.
From this perspective, our study complements that in \citet{DaughtonRKYABB2011}
by focusing on large-scale dynamics where a MHD description is applicable,
while neglecting kinetic physics that may be important at small scales.
In addition, because theories of anisotropic turbulence are much more
well developed for MHD than they are for collisionless plasmas, we
are able to apply well established diagnostics and make quantitative
comparisons with existing MHD turbulence theories.

This paper is organized as follows. Section \ref{sec:Simulation-Setup}
gives a detailed description of the simulation setup. The simulation
results are presented in Section \ref{sec:Simulation-Results}, which
is divided into two parts. Section \ref{sub:Reconnection-Rate} gives
an overall description of the time evolution and development of turbulence,
and makes comparison with corresponding 2D simulations with and without
plasmoids. Section \ref{sub:Characteristics-of-the-Turbulence} looks
more deeply into the characteristics of the fully developed turbulent
state, including power-law spectra of the inertial range, eddy anisotropy
with respect to local magnetic field, and comparisons with the Goldreich-Sridhar
theory of MHD turbulence \citep{GoldreichS1995,GoldreichS1997}. Finally,
we discuss the implications of our findings for large-scale astrophysical
reconnection and conclude in Section \ref{sec:Discussion-and-Conclusion}.

\section{Simulation Setup\label{sec:Simulation-Setup}}

\begin{figure}
\begin{centering}
\includegraphics[clip,scale=0.85]{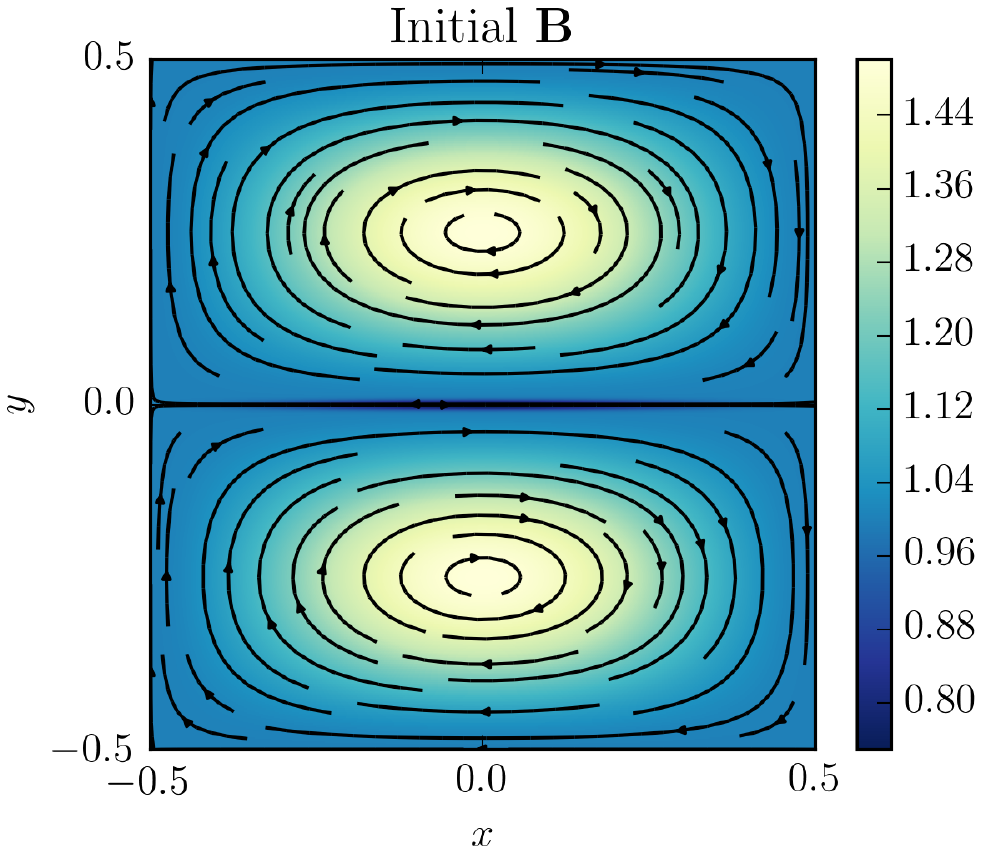}
\par\end{centering}

\caption{Initial magnetic field configuration. Black lines are stream lines
of the in-plane component, and color shading shows the out-of-plane
component $B_{z}$.\label{fig:Initial-magnetic-field}}
\end{figure}

The governing equations of our numerical model are the standard nondimensionalized
compressible, viscous, and resistive magnetohydrodynamics with an
adiabatic equation of state:

\begin{equation}
\partial_{t}\rho+\nabla\cdot\left(\rho\mathbf{v}\right)=0,\label{eq:1}
\end{equation}
\begin{equation}
\partial_{t}\left(\rho\mathbf{v}\right)+\nabla\cdot\left(\rho\mathbf{vv}\right)=-\nabla\left(p+\frac{B^{2}}{2}\right)+\nabla\cdot(\mathbf{BB})+\nu\nabla^{2}(\rho\mathbf{v}),\label{eq:2}
\end{equation}
\begin{equation}
\partial_{t}p+\nabla\cdot(p\mathbf{v})=-(\gamma-1)p\nabla\cdot\mathbf{v},\label{eq:3}
\end{equation}
\begin{equation}
\partial_{t}\boldsymbol{B}=\nabla\times\left(\mathbf{v}\times\boldsymbol{B}-\eta\mathbf{J}\right),\label{eq:4}
\end{equation}
\textcolor{black}{where standard notations are used. The electric
current density $\mathbf{J}$ is related to the magnetic field $\mathbf{B}$
via the relation $\mathbf{J}=\nabla\times\mathbf{B}$. }The numerical
algorithm is detailed in \citep{GuzdarDMHL1993}. Derivatives are
approximated by a five-point central finite difference scheme, with
a fourth-order numerical dissipation equivalent to up-wind finite
difference added to all equations for numerical stability. Time stepping
is calculated by a trapezoidal leapfrog scheme. Explicit dissipations
are employed through viscosity and resistivity.

We use a simulation setup that has been employed in previous studies
\citep{HuangB2010,HuangBS2011,HuangB2012}, where the attraction between
two coalescing magnetic flux tubes is the driver of magnetic reconnection.
The simulation box is \textcolor{black}{a 3D cube in the domain $(x,y,z)\in[-1/2,1/2]\times[-1/2,1/2]\times[-1/2,1/2]$}.
In normalized units, the initial magnetic field \textcolor{black}{is
given by $\mathbf{B}=B_{z}\mathbf{\hat{z}}+\mathbf{\hat{z}}\times\nabla\psi$,
where }$\psi=\tanh\left(y/h\right)\cos\left(\pi x\right)\sin\left(2\pi y\right)/2\pi$,
and $B_{z}$ is non-uniform such that the initial configuration is
approximately force-balanced (Figure \ref{fig:Initial-magnetic-field}).
The parameter $h$, which is set to $1/300$ for this study, determines
the initial width of the current layer between the flux tubes. In
the upstream region of the current layer, the reconnecting component
$B_{x}$ and the guide field $B_{z}$ are both approximately unity.
The initial plasma density and temperature are both uniform, with
$\rho=1$ and $T=1$ in normalized units. The viscosity $\mu$ and
resistivity $\eta$ are both set to $5\times10^{-6}$, which give
a Lundquist number $S\equiv V_{A}L/\eta=2\times10^{5}$ and a magnetic
Prandtl number $Pr_{m}\equiv\mu/\eta=1$. {The heat
capacity ratio $\gamma=5/3$ is assumed.} Perfectly conducting and
free slipping boundary conditions are imposed along both $x$ and
$y$ directions, and periodic boundary conditions along $z$. The
simulation mesh size is $N_{x}\times N_{y}\times N_{z}=2000\times1000\times2000$,
where the grid sizes are uniform along both $x$ and $z$ directions,
and packed along the $y$ direction around the midplane to better
resolve the reconnection layer. The grid size along $y$ near the
midplane ($y=0$) is $\Delta y=10^{-4}$, which gradually increases
away from the midplane and reaches $\Delta y=0.005$ near the boundary.
The resolution on the $x-y$ plane is based on our extensive experience
from a 2D scaling study \citep{HuangB2010}, where simulations with
different resolutions have been performed. In this study, we have
also performed the same simulation with $N_{z}=1000$ and $2000$
and found no significant difference. To trigger the plasmoid instability,
the initial velocity is seeded with a random noise of amplitude $10^{-3}$.
This is in contrast with the situation in particle-in-cell codes,
where noise associated with a finite number of particles is enough
to trigger the plasmoid instability.

\section{Simulation Results\label{sec:Simulation-Results}}

\subsection{Time Evolution and Reconnection Rate\label{sub:Reconnection-Rate}}

\begin{figure}
\begin{centering}
\includegraphics[clip,scale=0.82]{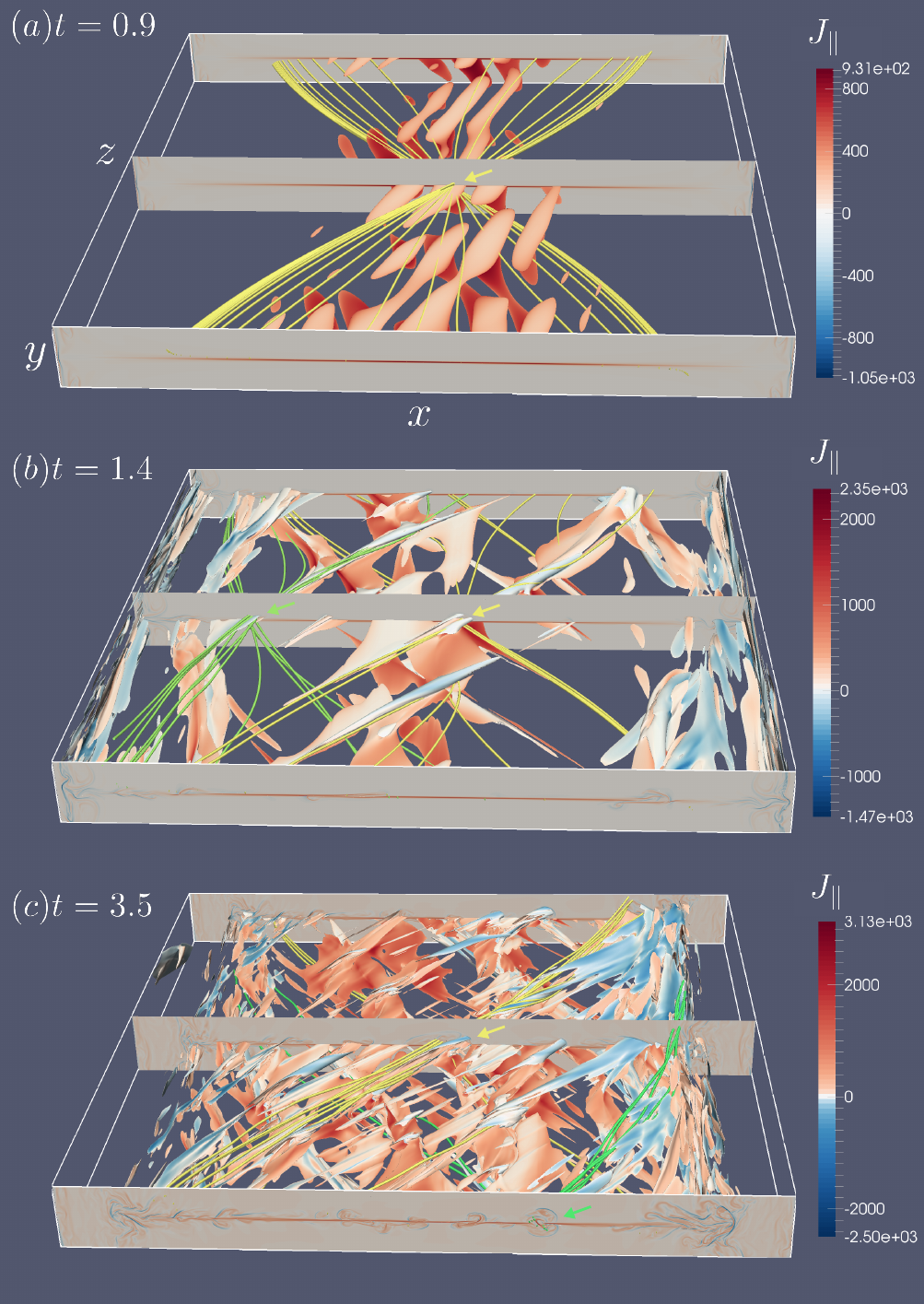}
\par\end{centering}

\caption{Snapshots of the 3D simulation at three representative times. Color
shading shows the component of the electric current parallel to the
magnetic field $J_{\parallel}\equiv\mathbf{J}\cdot\mathbf{\hat{b}}$
on three $x-y$ slices, as well as on isosurfaces of the fluctuating
part of the magnetic energy $\tilde{E_{m}}=|\tilde{\mathbf{B}}|^{2}/2$.
These snapshots also show samples of magnetic field lines, where field
lines with the same color are originated from a selected small region
as indicated by an arrow of the same color. These plots show the entire
$x$ and $z$ dimensions of the simulation box, but only the region
$-0.05\le y\le0.05$ along the $y$ direction.\label{fig:Snapshots} }
\end{figure}
\begin{figure}
\begin{centering}
\includegraphics[clip,scale=0.65]{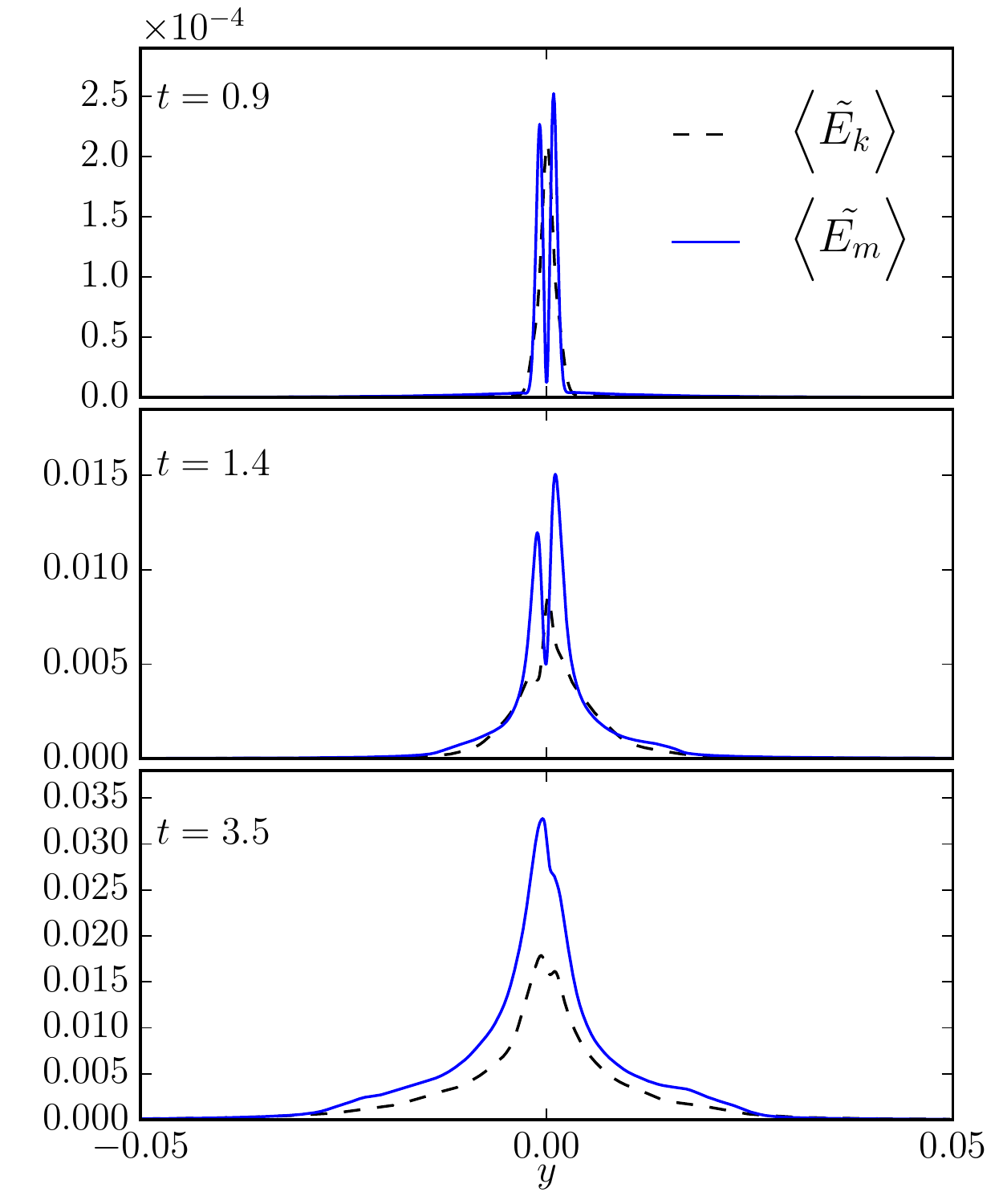}
\par\end{centering}

\caption{One-dimensional profiles (along $y$) of the averaged kinetic energy
fluctuation $\left\langle \tilde{E_{k}}\right\rangle $ and magnetic
energy fluctuation $\left\langle \tilde{E_{m}}\right\rangle $, corresponding
to the three snapshots of Figure \ref{fig:Snapshots}. Here the averaging
is carried out over the central part of the reconnection layer within
the range $-0.25\le x\le0.25$ and the entire $z$ direction.\label{fig:EkEm1d}}
\end{figure}

{Our primary interest of this study is to understand
how small-scale turbulence affects reconnection of a large-scale ``mean''
magnetic field. To separate the large-scale mean field from small-scale
fluctuations, the former must be determined through an averaging procedure
(or coarse-graining). Two commonly used definitions for the mean field
are ensemble average and time average. The former is obtained via
averaging over the ensemble of all turbulent realizations (e.g., from
different initial random noise) of the same setting, whereas the latter
is obtained by averaging over an appropriate period of time. If we
further assume ergodicity (see, e.g. the discussion in \citet{Frisch1995}),
ensemble average and time average are nearly equivalent. In practice,
a time average is often adopted, because an ensemble average requires,
by definition, many different realizations of the same setup, which
can be prohibitively expensive. However, because the system under
consideration has a translational symmetry along the $z$ direction,
i.e. $z$-dependence only arises as a result of instabilities, after
averaging over the entire ensemble, the mean field must be independent
of $z$. Therefore, instead of using a time average, we adopt the
convention of using the average of a physical variable $f$ over the
entire $z$ direction as the mean field $\bar{f}$, which is taken
as a proxy for the ensemble average. Once the mean field is determined,
the fluctuation is obtained by the remaining part $\tilde{f}\equiv f-\bar{f}$
. This procedure ensures that the mean magnetic field $\bar{\mathbf{B}}$
is independent of $z$, and consequently we may calculate the 3D reconnection
rate in terms of $\bar{\mathbf{B}}$ in the same way as in 2D cases.
We should note, however, that this procedure is not applicable for
more general situations when the initial condition is dependent on
all three coordinates. In that case, an appropriate definition for
the mean field will be a time average over a period of time sufficiently
longer than typical turbulence eddy turnover time, but shorter than
evolution time scales of large scale field. The resulting mean field
in general will depend on all three coordinates. Nevertheless, a general
definition of reconnection rate in full 3D configurations remains
a topic of debate, which is beyond the scope of this work. See, e.g.
the discussion in \citep{HuangBB2014,DaughtonNKRL2014,WyperH2015}
and the references therein. }

We begin by examining the time evolution and development of turbulence
in the reconnection layer. Figure \ref{fig:Snapshots} shows three
representative snapshots of the reconnection layer, where color shading
shows the component of the electric current parallel to the magnetic
field $J_{\parallel}\equiv\mathbf{J}\cdot\mathbf{\hat{b}}$ on three
$x-y$ slices, as well as on isosurfaces of the fluctuating part of
the magnetic energy $\tilde{E_{m}}=|\tilde{\mathbf{B}}|^{2}/2$. These
snapshots also show samples of magnetic field lines, where field lines
with the same color originate from a selected small region as indicated
by an arrow of the same color. Here the isosurfaces in each snapshot
correspond to a single value of $\tilde{E_{m}}$; they are employed
as a means to visualize the development of complex structures as the
instabilities evolve. The color shaded parallel electric current $J_{\parallel}\equiv\mathbf{J}\cdot\mathbf{\hat{b}}$
is employed as a proxy for showing where non-ideal effects are concentrated.
Panel (a) shows an early phase when the plasmoid instabilities are
developing, at $t=0.9$. It shows that magnetic fluctuations initially
develop preferentially at oblique angles, at locations slightly away
from the midplane. At this time the Sweet-Parker current sheet is
still largely unperturbed. Panel (b) shows a snapshot at $t=1.4$,
when the instabilities have further developed, and some coherent structures
start to become visible on the $x-y$ slices of the $J_{\parallel}$
profiles. Panel (c) shows a snapshot at $t=3.5$, when the instabilities
have developed into a fully turbulent state. At this time, the isosurfaces
of $\tilde{E_{m}}$ form complicated structures, which appear to align
preferentially with magnetic field lines. This is an important feature
that we will come back to at later discussion. The $x-y$ slices of
$J_{\parallel}$ also show blob-like structures which give the impression
that they may be cross sections of magnetic flux ropes. However, tracing
field lines from the blobs shows that not to be the case. The two
sets of field lines (indicated by yellow and green colors, respectively)
in panel (c) both originate from a blob-like structure, but the field
lines clearly show the influence of the global magnetic shear across
the reconnection layer. Each set of field lines roughly separates
into two bundles, one approximately follows the magnetic field above
the reconnection layer, while the other approximately follows the
magnetic field below the layer. In these two sets of field lines,
however, some neighboring field lines are found to wrap around each
other over a certain distance, which may be loosely interpreted as
indicative of flux ropes. 

To see how the turbulent region broadens as the instabilities develop,
we calculate the one-dimensional (1D) profiles of averaged kinetic
energy fluctuation $\left\langle \tilde{E_{k}}\right\rangle $ and
magnetic energy fluctuation $\left\langle \tilde{E_{m}}\right\rangle $
along the $y$ direction. Here the averaging is carried out over the
central part of the reconnection layer within the range $-0.25\le x\le0.25$
and the entire $z$ direction. Because we employ a compressible MHD
model, the kinetic energy fluctuation is defined through a new variable
$\mathbf{w}\equiv\sqrt{\rho}\mathbf{u}$ \citep{KidaO1992} such that
the kinetic energy density $E_{k}=w^{2}/2$ is a quadratic form of
$w$ and the fluctuation part of kinetic energy density is defined
as $\tilde{E_{k}}\equiv\left|\tilde{\mathbf{w}}\right|^{2}/2$. Figure
\ref{fig:EkEm1d} shows the resulting 1D profiles corresponding to
the three snapshots of Figure \ref{fig:Snapshots}. The top panel
at $t=0.9$ shows that $\left\langle \tilde{E_{m}}\right\rangle $
has two peaks, corresponding to the regions of magnetic energy fluctuations
growing at oblique angles in Figure \ref{fig:Snapshots} (a). In contrast,
the kinetic energy fluctuation $\left\langle \tilde{E_{k}}\right\rangle $
peaks at the center around $y=0$. These energy fluctuations are localized
to a narrow layer within the Sweet-Parker layer. As the instabilities
develop, the energy fluctuations gradually spread out to a broader
region, shown in the middle panel at $t=1.4$. At this time the magnetic
energy fluctuation profile still exhibits two peaks. The bottom panel
of Figure \ref{fig:EkEm1d} shows the fully developed state at $t=3.5$,
when the turbulent region becomes even broader. At this time the double-peak
structure of $\left\langle \tilde{E_{m}}\right\rangle $ has almost
disappeared. Magnetic fluctuations carry more energy than kinetic
fluctuations, and approximately $70\%$ of the total energy fluctuation
is contained in the range $-0.01\le y\le0.01$.

\begin{figure}
\begin{centering}
\includegraphics[scale=0.65]{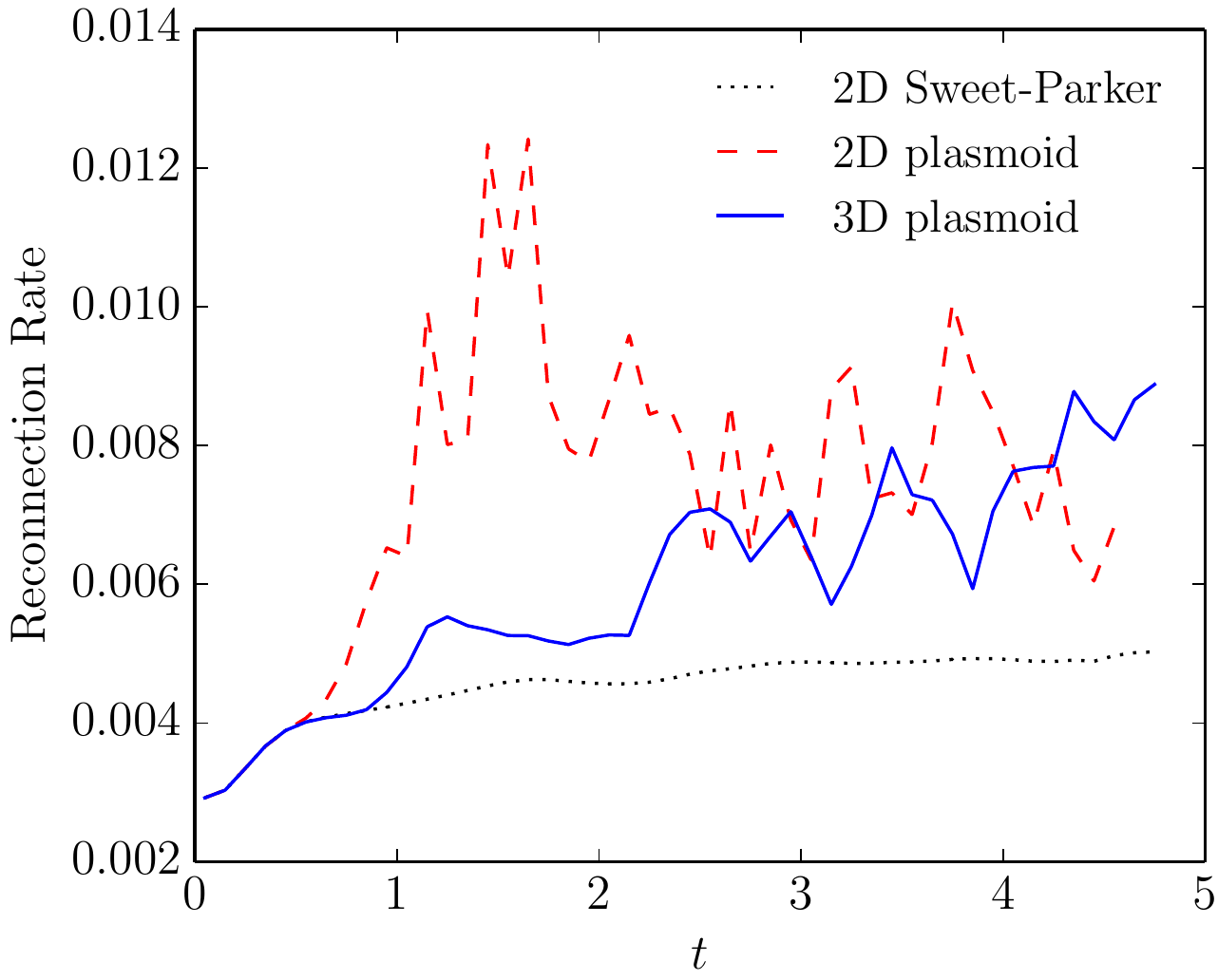}
\par\end{centering}

\caption{Time histories of reconnection rates for the three runs. \label{fig:reconnection-rate}}
\end{figure}

As mentioned in the Introduction, it has been established that the
plasmoid instability leads to fast reconnection in 2D resistive MHD
with the reconnection rate nearly independent of the resistivity.
An important question is how 3D effects affect the overall reconnection
rate. To facilitate a comparison, we carry out two additional 2D simulations
with the same setting. The first one without initial random noise
realizes Sweet-Parker reconnection, whereas the second one with an
initial random noise of the same amplitude as the 3D case results
in 2D plasmoid-dominated reconnection. We measure the reconnection
rate in the following way. First we use the mean field $\bar{\mathbf{B}}$
(which is simply $\mathbf{B}$ in 2D cases) to calculate the reconnected
magnetic flux as the maximum of the flux function $\psi(x)\equiv\int_{-1/2}^{x}\bar{B_{y}}(x')dx'$
along the midplane ($y=0$). Its time derivative
$d\psi/dt$ then gives the instantaneous reconnection rate. Figure
\ref{fig:reconnection-rate} shows the time histories of the reconnection
rates for the three cases. The three curves coincide with each other
at the beginning. The plasmoid instability sets in at an earlier time
$t\simeq0.5$ in the 2D simulation, and the reconnection rate rapidly
increases and reaches the peak value $d\psi/dt\simeq0.012$ at $t\simeq1.5$,
then decreases to a quasi-steady plateau that fluctuates around $d\psi/dt\simeq0.008$.
In contrast, the plasmoid instability in the 3D simulation sets in
at a relatively later time $t\simeq0.9$. The reconnection rate does
not rapidly increase to a peak then decreases as in the 2D case. Instead,
it gradually increases and reaches a quasi-steady value that is comparable
to the 2D counterpart and appears to be slightly increasing over time.
Not surprisingly, the Sweet-Parker run gives the slowest reconnection
rate, which approximately reaches $d\psi/dt\simeq0.004$ after the
initial current sheet thinning period, and slightly increases over
the entire simulation time and reaches $d\psi/dt\simeq0.005$ at the
end of the simulation. The slowly increasing reconnection rates in
both Sweet-Parker reconnection and 3D turbulent reconnection may be
attributed to gradual evolution of the global configuration. This
simulation setup does not allow a true steady state as the two merging
flux tubes that drive the reconnection process shrink in size as time
passes. The reconnection rates of the 2D and 3D simulations with plasmoid
instabilities are less than a factor of two higher than the Sweet-Parker
reconnection rate. This relatively low enhancement of reconnection
rate is due to the modest value of Lundquist number $S=2\times10^{5}$
we are able to do in 3D. This modest Lundquist number also makes the
Sweet-Parker reconnection realizable when the initial condition is
not seeded with noise. For significantly higher Lundquist numbers
that we have realized in previous 2D studies, the current sheet becomes
so fragile that plasmoid instabilities set in even if the system is
not seeded with an initial noise. 

Our results indicate that reconnection rates in 2D and 3D are comparable
once the system reaches a quasi-steady state. Interestingly, plasmoid
instabilities appear to set in earlier in the 2D simulation than in
the 3D run. This may be attributed to the fact that the fastest growing
2D modes are faster than oblique modes \citep{BaalrudBH2012}. Therefore,
plasmoid instabilities in the 2D simulation, which effectively start
from an initial noise that is uniform along the $z$ direction, undergo
a more rapid growth than in the 3D simulation.  Furthermore, the
reconnection rate in 2D plasmoid-dominated reconnection reaches a
higher value before settling down to a quasi-steady rate. Although
we have not reported this feature in our earlier papers, it appears
to be a general trend in our previous 2D studies as well. The reason
for this feature is as follows: during the early phase when the Sweet-Parker
current sheet goes through rapid fractal-like fragmentation \citep{HuangB2013},
there are usually more plasmoids than at a later time when the system
reaches a quasi-steady state in which the formation of new plasmoids
is balanced by loss of plasmoids due to advection and coalescence
\citep{HuangB2012}. Therefore during the early phase the secondary
current sheets tend to be shorter and narrower compared to the secondary
current sheets at a later time, hence the reconnection rate is higher.
In comparison, current sheet fragmentation in the 3D simulation appears
to set in at a more gentle pace. 

\begin{figure}
\begin{centering}
\includegraphics[clip,scale=0.8]{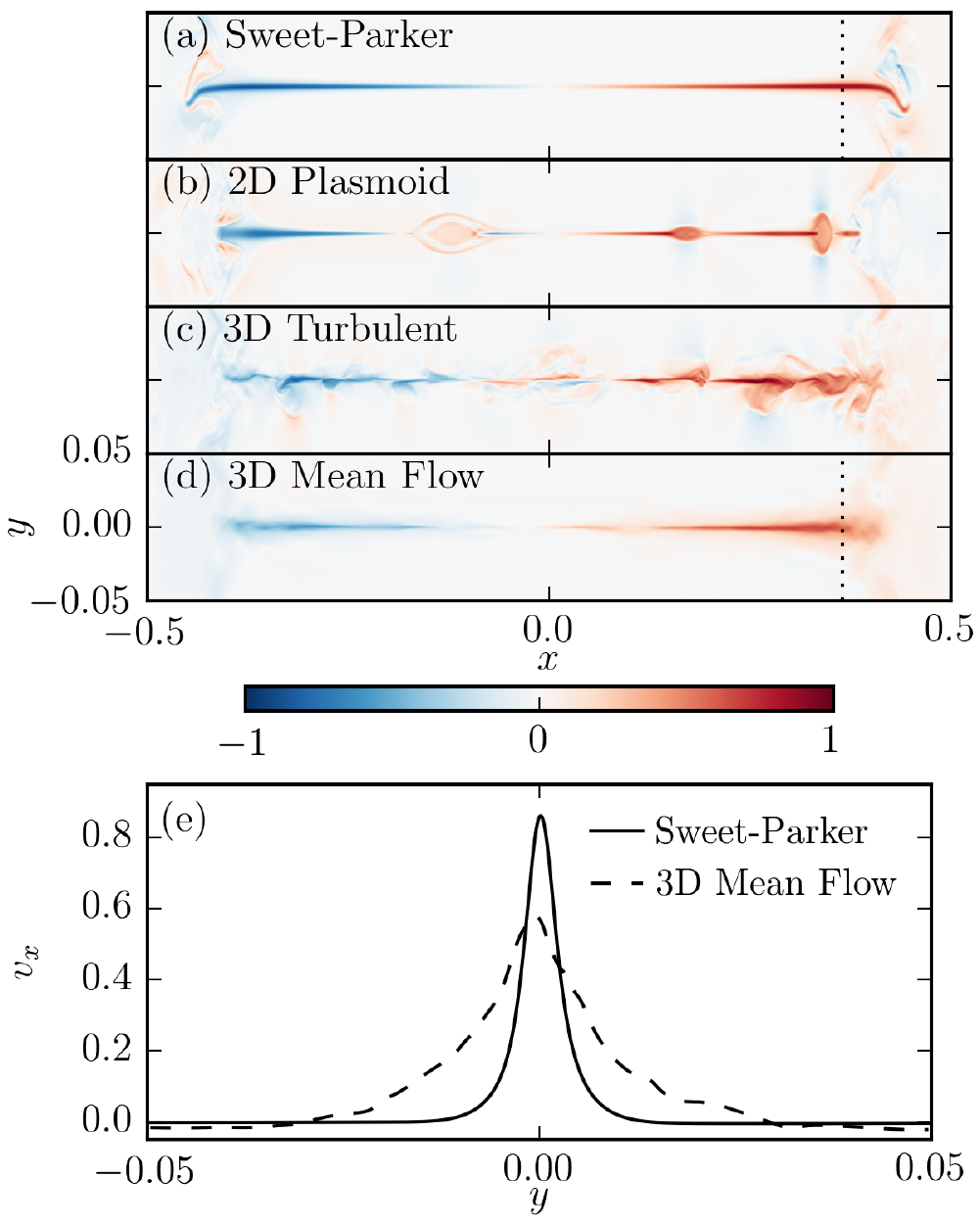}
\par\end{centering}

\caption{Comparison of outflow ($v_{x}$) profiles. (a) 2D Sweet-Parker reconnection
at $t=4.2$; (b) 2D plasmoid-dominated reconnection at $t=2.6$; (c)
slice of 3D turbulent reconnection at $z=0$; and (d) the mean field
$\bar{v_{x}}$ of the 3D turbulent reconnection at $t=3.5$. The reconnected
fluxes at the selected snapshots are approximately the same to make
a fair comparison. Panel (e) shows one-dimensional cuts of the outflow
profiles along the dotted lines in panel (a) and panel (d); the 3D
mean flow is considerably broader than the Sweet-Parker outflow.\label{fig:Comparison-of-outflow}}
\end{figure}
\begin{figure}
\begin{centering}
\includegraphics[clip,scale=0.65]{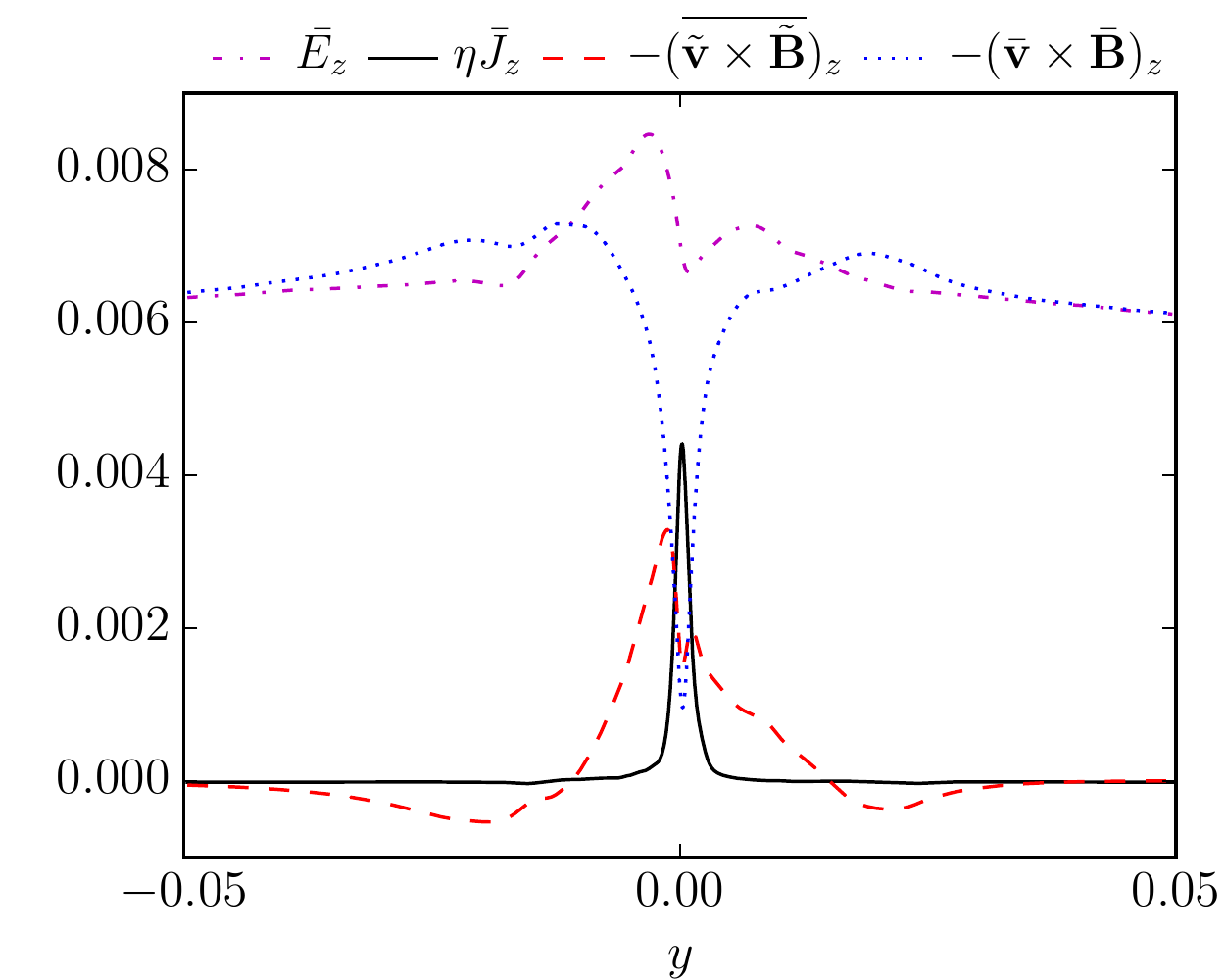}
\par\end{centering}

\caption{Decomposition of the out-of-plane mean electric field $\bar{E_{z}}$,
taken at $t=3.5$ and averaged over the range from $x=-0.05$ to $x=0.05$.\label{fig:Decomposition-Ohm}}
\end{figure}

Magnetic reconnection converts magnetic energy into plasma kinetic
energy by turning plasma from upstream regions into Alfv\'enic outflow
jets, which also transport plasma from upstream regions to downstream
regions. It is therefore illuminating to compare the outflow profile
$v_{x}$ from three simulations, shown in Figure \ref{fig:Comparison-of-outflow}.
In order to make a fair comparison, these snapshots are taken at times
when the reconnected magnetic fluxes are approximately the same for
the three cases. Panel (a) shows the laminar, bi-directional Sweet-Parker
outflow jets. In panel (b), some well-defined plasmoids are clearly
visible. Panel (c) shows a $x-y$ slice at $z=0$ for the 3D run,
at $t=3.5$ when turbulent reconnection is fully developed. Here some
coherent structures similar to the plasmoids in panel (b) are still
visible, but the structures are less regular. However, as we average
over the $z$ direction to obtain the mean flow $\bar{v_{x}}$, shown
in panel (d), it appears similar to a ``blurred'' Sweet-Parker outflow
profile as in panel (a). One-dimensional cuts of the outflow profiles
at the exhaust regions along the dotted lines in panel (a) and panel
(c), shown in panel (e), further reinforce this impression. The mean
outflow jet of 3D turbulent reconnection is substantially broader
than the Sweet-Parker outflow. The ratio between the areas below the
two curves, which measure the total outflow fluxes, is consistent
with the enhancement of reconnection rate obtained from the magnetic
flux measurement. 

Because reconnection of the mean magnetic field must be driven by
an out-of-plane mean electric field $\bar{E_{z}}$, it is useful to
decompose $\bar{E_{z}}$ into contributions from various terms:
\begin{equation}
\bar{E_{z}}=-(\bar{\mathbf{v}}\times\bar{\mathbf{B}})_{z}-\overline{(\tilde{\mathbf{v}}\times\tilde{\mathbf{B}})}_{z}+\eta\bar{J_{z}},\label{eq:ohm}
\end{equation}
where $-\overline{(\tilde{\mathbf{v}}\times\tilde{\mathbf{B}})}_{z}$
is the turbulent emf (electromotive force). Figure \ref{fig:Decomposition-Ohm}
shows an 1D cut of the decomposition along the inflow ($y)$ direction.
Here each term in Eq. (\ref{eq:ohm}) has been averaged over the range
$-0.05\le x\le0.05$ to further reduce the fluctuations. These curves
show that in the outer region, $\bar{E_{z}}$ is mostly balanced by
the $-(\bar{\mathbf{v}}\times\bar{\mathbf{B}})_{z}$ term, whereas
in the inner region, $\bar{E_{z}}$ is balanced by the resistive term
$\eta\bar{J}_{z}$ and the turbulent emf term $-\overline{(\tilde{\mathbf{v}}\times\tilde{\mathbf{B}})}_{z}$.
Note that $\eta\bar{J}_{z}$ is only significant in the innermost
narrow region, whereas the turbulent emf covers a much wider region.
 Interestingly, we find that the current sheet width of mean field
$\bar{J_{z}}$ is approximately the same as the corresponding Sweet-Parker
current sheet width, which also manifests in the fact that $\eta\bar{J_{z}}\simeq0.004$
at the center is approximately the same as the Sweet-Parker reconnection
rate. Whether this is a general feature or simply a coincidence is
not clear at this point, and should be examined in the future with
simulations of higher Lundquist numbers. Nonetheless, our result indicates
that although turbulence is effective in broadening the mean field
outflow jets (relative to Sweet-Parker reconnection), it is less effective
in broadening the mean field current sheet, if at all. This finding
suggests that the effect of turbulence does not lead to an enhanced
anomalous resistivity, which requires that $-\overline{(\tilde{\mathbf{v}}\times\tilde{\mathbf{B}})}_{z}$
be proportional to $\bar{J_{z}}$, but whether it may be describable
in terms of hyper-resistivity \citep{BhattacharjeeH1986,VanBallegooijenC2008}
is left to future work. 

From these analyses, we conclude that 3D turbulence enhances reconnection
rate by effectively broadening the reconnection layer and outflow
jets, where the turbulent emf provides important contribution to the
reconnecting electric field. This picture of 3D turbulent reconnection
is quite different from 2D plasmoid-dominated reconnection. In the
2D case, the reconnection rate is sped up because plasmoid instabilities
cause secondary current sheets to become shorter and narrower. Reconnection
takes place at multiple sites with enhanced local reconnection rates,
which transport magnetic flux and plasma into plasmoids. Finally,
reconnected magnetic flux and plasma are transported to the downstream
region when plasmoids are ejected from the reconnection layer, typically
at a Alfv\'enic time scale \citep{GunterYLBH2015}.

\subsection{Characteristics of the Self-Sustained Turbulence\label{sub:Characteristics-of-the-Turbulence}}

\begin{figure*}
\begin{centering}
\includegraphics[clip,scale=0.8]{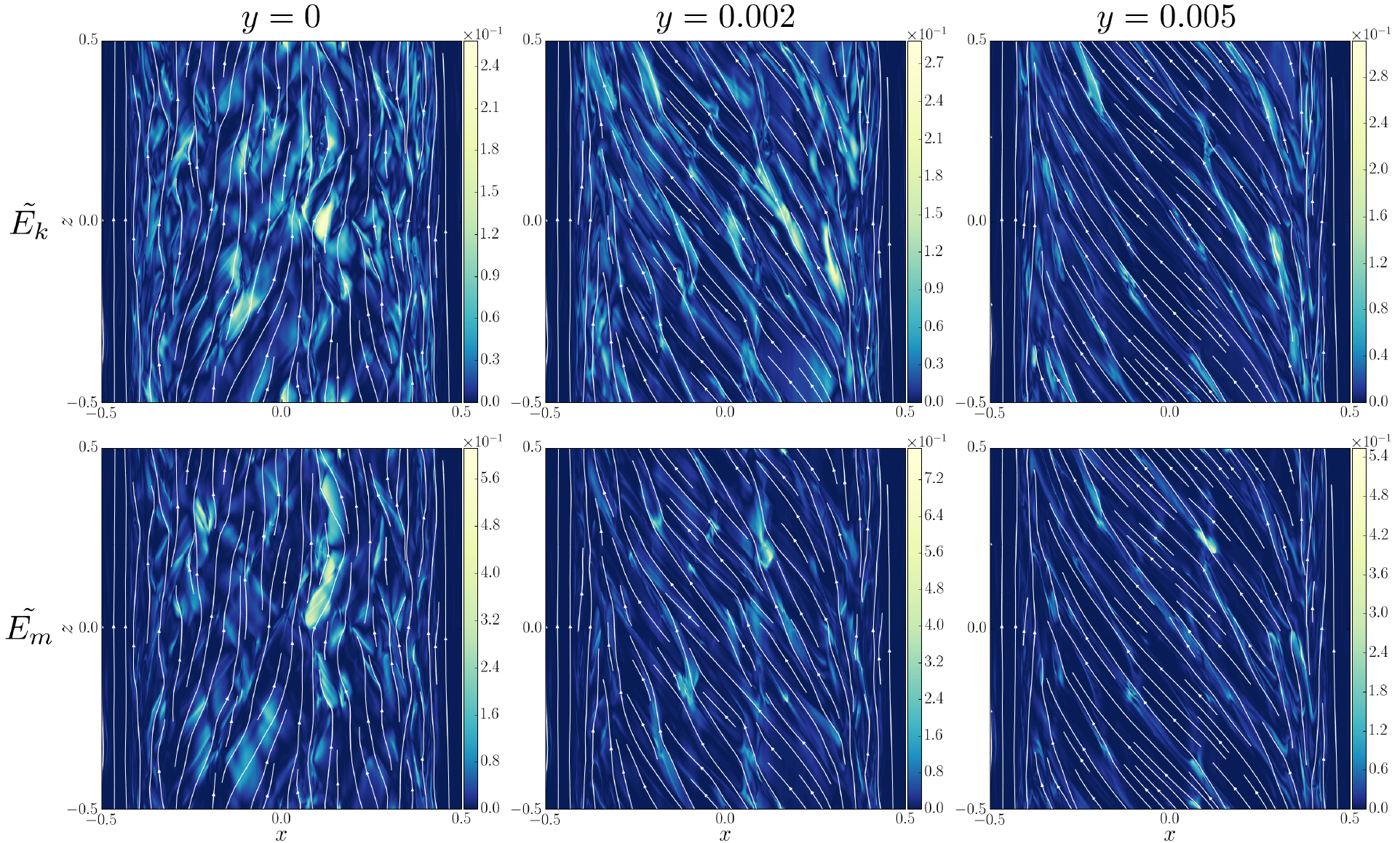}
\par\end{centering}

\caption{Energy fluctuations $\tilde{E_{k}}$ (first row) and $\tilde{E_{m}}$
(second row) at three different $x-z$ slices, overlaid with stream
lines of the in-plane component of the magnetic field. These snapshots
are taken at $t=3.5$, when turbulence in the reconnection layer has
fully developed. The energy fluctuations form cigar-shaped eddies
elongated along the direction of local magnetic field, which is one
of the hallmarks of MHD turbulence.\label{fig:Energy-fluctuations}}
\end{figure*}

\begin{figure}
\begin{centering}
\includegraphics[clip,scale=0.65]{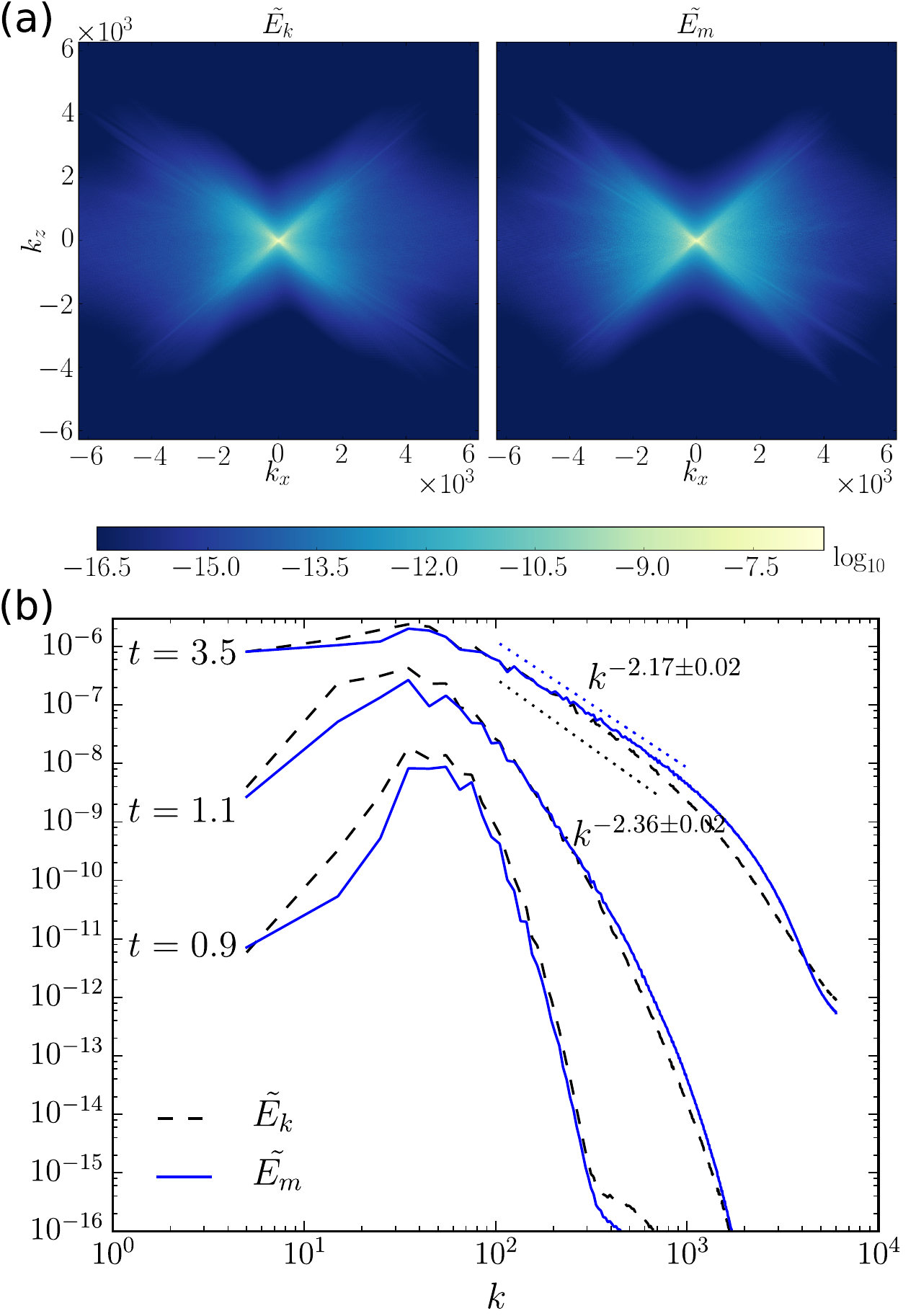}
\par\end{centering}

\caption{(a) Kinetic and magnetic energy spectra in Fourier space, integrated
over the range from $y=-0.05$ to $0.05$ at $t=3.5$. Both spectra
are qualitatively similar and highly anisotropic. Fourier modes are
mostly excited within the region $\left|k_{x}\right|\gtrsim\left|k_{z}\right|$,
as dictated by the resonant condition $\mathbf{k}\cdot\mathbf{B}\simeq0$.
(b) One-dimensional spectra at $t=0.9$, $t=1.1$, and $t=3.5$ obtained
by integrating over the azimuthal direction on the $k_{x}-k_{z}$
plane.\label{fig:spectra}}

\end{figure}

\begin{figure}
\begin{centering}
\includegraphics[scale=0.9]{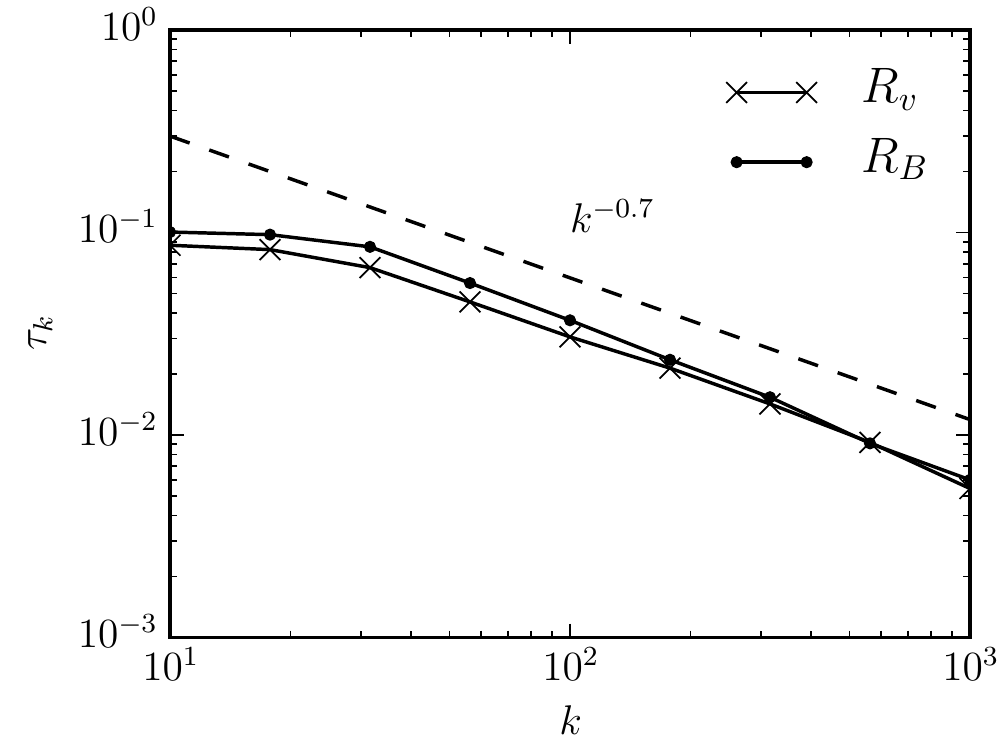}
\par\end{centering}

\caption{Eddy turnover times estimated by scale-dependent e-folding decay times
of autocorrelation functions $R_{v}(k,\tau)$ and $R_{B}(k,\tau)$.\label{fig:Eddy-turnover-times}}
\end{figure}
\begin{figure*}
\begin{centering}
\includegraphics[clip,scale=0.9]{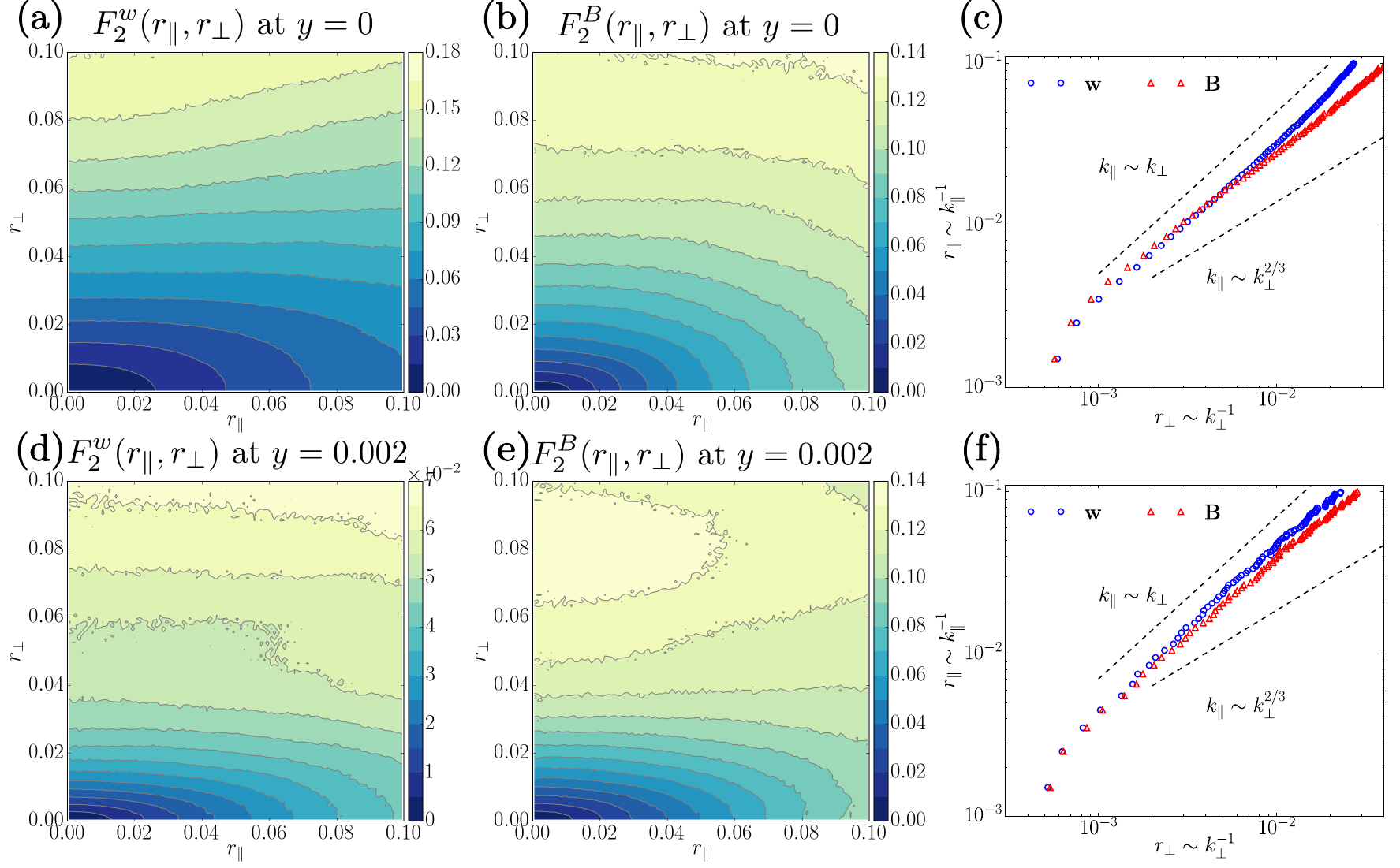}
\par\end{centering}

\caption{Two-point structure functions of the fully developed turbulence at
$t=3.5$. (a) $F_{2}^{w}(r_{\parallel},r_{\perp})$ and (b) $F_{2}^{B}(r_{\parallel},r_{\perp})$
at $y=0$. (c) Relationships between semi-major axis $r_{\parallel}\sim k_{\parallel}^{-1}$
and semi-minor axis $r_{\perp}\sim k_{\perp}^{-1}$ of contours in
(a) and (b), which measure the scale dependency of turbulent eddy
anisotropy. The two dashed lines represent the relations $k_{\parallel}\sim k_{\perp}$
(scale-independent) and $k_{\parallel}\sim k_{\perp}^{2/3}$ (GS theory),
for reference. Panels (d) -- (f) are the same as panels (a) -- (c),
except at $y=0.002$. \label{fig:Two-point-structure-functions} }
\end{figure*}

Now we examine further the characteristics of the self-sustained turbulent
state within the reconnection layer and make comparison with the Goldreich
\& Sridhar theory of incompressible MHD turbulence \citep{GoldreichS1995,GoldreichS1997}.
In the following discussion we present detailed diagnostics of the
turbulent state at $t=3.5$, but the characteristics we find here
are generally valid during the period when the turbulence has fully
developed, after $t=3.0$. 

The first feature we notice is that the turbulent state is highly
inhomogeneous --- the mean magnetic field is strongly sheared over
a short distance from $y=-0.002$ to $y=0.002$ (i.e. the mean field
current sheet width, see Figure \ref{fig:Decomposition-Ohm}), and
the turbulence is embedded in a mean flow which is also highly sheared
(Figure \ref{fig:Comparison-of-outflow} (d)). Figure \ref{fig:Energy-fluctuations}
shows the kinetic and magnetic energy fluctuations, overlaid with
streamlines of the in-plane component of the magnetic field, on three
representative $x-z$ slices at the midplane $(y=0)$, the edge of
the mean field current sheet ($y=0.002$), and a further outer region
($y=0.005$). These energy fluctuations form cigar-shaped eddies elongated
along the direction of local magnetic field, which is one of the hallmarks
of MHD turbulence. It is observed that the eddies become more elongated
away from the midplane, likely because the magnetic field becomes
less sheared in those regions. {Fluctuations are strongest
near the midplane, where the dominant magnetic field is guide field
component $B_{z}\simeq1$, and the dominant flow component is the
outflow jet $v_{x}$, which can be up to the Alfv\'en speed $V_{A}\simeq1$.
At the midplane, the root-mean-square (RMS) values of magnetic field
fluctuations $\tilde{B_{x}}$, $\tilde{B_{y}}$, and $\tilde{B_{z}}$
are approximately $0.22$, $0.05$, and $0.07$, whereas the RMS values
of $\tilde{v_{x}}$, $\tilde{v_{y}}$, and $\tilde{v_{z}}$ are approximately
0.16, 0.02, and 0.05, respectively. However, as can be seen from Figure
\ref{fig:Energy-fluctuations}, fluctuations $\tilde{B}$ and $\tilde{v}$
can locally be as high as $O(1)$ of the guide field and the Alfv\'en
speed $V_{A}$.}

{A common feature of turbulent systems is the formation
of an extended power-law inertial range in energy spectra through
cascade, which is what we will examine next. }However, because our
turbulent system is spatially inhomogeneous, a proper measurement
of energy spectra is a challenge. The procedure we adopt is as follows.
Because our primary interest is the turbulence \emph{within} the reconnection
layer, we multiply the fluctuating fields $\tilde{\mathbf{B}}$ and
$\tilde{\mathbf{w}}$ by a $C^{\infty}$ Planck-taper window function
\citep{McKechanRS2010} which equals unity within the range $-0.2\le x\le0.2$
and tapers off smoothly to zero over the ranges where $0.2\le\left|x\right|\le0.4$.
This step effectively filters out the energy fluctuations in the downstream
regions which may be of different characteristics. We then calculate
the discrete Fourier energy spectra of $\tilde{E_{k}}$ and $\tilde{E_{m}}$
using the ``windowed'' variables on each $x-z$ plane to obtain
2D energy spectra in terms of wave numbers $k_{x}$ and $k_{z}$.
Finally, we integrate the 2D energy spectra over $-0.05\le y\le0.05$,
which is the direction of the strongest inhomogeneity. The resulting
2D spectra of $\tilde{E_{k}}$ and $\tilde{E_{m}}$ as functions of
$k_{x}$ and $k_{z}$ are shown in Figure \ref{fig:spectra} (a).
It can be seen that the spectra of $\tilde{E_{k}}$ and $\tilde{E_{m}}$
are qualitatively similar, and both lie predominantly in the regions
where $\left|k_{x}\right|\gtrsim\left|k_{z}\right|$. This is a consequence
of the fact that the energy fluctuations tend to align preferentially
with local magnetic field, i.e. $\mathbf{k}\cdot\mathbf{B}\simeq0$,
whereas $B_{z}\simeq1$ and $B_{x}$ varies approximately from $-1$
to $1$ in the reconnection layer. Although these energy spectra are
anisotropic, we may still calculate 1D spectra by integrating over
the azimuthal direction on the $k_{x}-k_{z}$ plane. Figure \ref{fig:spectra}
(b) shows the resulting 1D spectra of $\tilde{E_{k}}$ and $\tilde{E_{m}}$,
both of which exhibit an extended inertial range with a power-law
spectrum $\sim k^{-\alpha}$. The kinetic energy spectrum is slightly
steeper than the magnetic energy spectrum in the inertial range. During
the period from $t=3.0$ to $4.8$ when the turbulence has fully developed,
the power-law index $\alpha$ is typically within the range $2.1<\alpha<2.3$
for $\tilde{E_{m}}$, and $2.3<\alpha<2.5$ for $\tilde{E_{k}}$.
One-dimensional power-law spectra with similar power indices are also
obtained by integrating over either $k_{x}$ or $k_{z}$ directions.
Figure \ref{fig:spectra} (b) also shows the 1D spectra at $t=0.9$
and $t=1.1$ when the instabilities are still in early stages of development.
It shows that the instabilities initially inject energy predominantly
in intermediate wave numbers $k\sim30$ -- $100$, which gradually
cascades down to smaller scales (high-$k$ modes) and form the inertial
range, while at the same time also develops coherent structures at
larger scales, manifested as low-$k$ modes in the spectra. 

{Eddy turnover times at different length scales can
be estimate by calculating scale-dependent autocorrelation time of
relevant variables. Specifically, we calculate $R_{v}(k,\tau)\equiv\int d^{3}x\tilde{\mathbf{v}}_{k}^{>}(\mathbf{x},t)\cdot\tilde{\mathbf{v}}_{k}^{>}(\mathbf{x},t+\tau)$
and $R_{B}(k,\tau)\equiv\int d^{3}x\tilde{\mathbf{B}}_{k}^{>}(\mathbf{x},t)\cdot\tilde{\mathbf{B}}_{k}^{>}(\mathbf{x},t+\tau)$,
where $\tilde{\mathbf{B}}_{k}^{>}$ and $\tilde{\mathbf{v}}_{k}^{>}$
denote high-pass filtered $\tilde{\mathbf{B}}$ and $\tilde{\mathbf{v}}$
with wavenumbers $k_{x}$ and $k_{z}$ satisfying the condition $k_{x}^{2}+k_{z}^{2}>k^{2}$,
and the integrals are carried out over the region $-0.05\le y\le0.05$.
It is found that $R_{v}(k,\tau)$ and $R_{B}(k,\tau)$ both approximately
decay exponentially with $\tau$ over a broad range of length scales
set by the wavenumber $k$. This allows us to calculate scale-dependent
e-folding decay times as proxies for eddy turnover times, and the
results are shown in Figure \ref{fig:Eddy-turnover-times}. We find
that both autocorrelation functions $R_{v}(k,\tau)$ and $R_{B}(k,\tau)$
give similar estimates for the eddy turnover time $\tau_{k}$, which
approximately follow a $\tau_{k}\sim k^{-0.7}$ power law for $k\gtrsim30$,
while $\tau_{k}\simeq0.1$ at the largest scales. Once the characteristic
time scales have been established, we may further examine the validity
of our convention of using an average over $z$ for the mean field
by comparing with results from using a time average for the mean field.
Here the interval for time averaging has to be sufficiently longer
than the eddy turnover times at all scales, while sufficiently shorter
than the evolution time scales of large-scale magnetic field. We have
calculated the fluctuation energy spectra using an interval $\Delta t=0.275$
for time averaging, and indeed the results are similar to that in
Figure \ref{fig:spectra}.}

To further investigate the alignment of energy fluctuations with local
magnetic field, we calculate two-point structure functions in terms
of parallel displacement $r_{\parallel}$ and perpendicular displacement
$r_{\perp}$ with respect to local magnetic field, i.e. $F_{2}^{w}(r_{\parallel},r_{\perp})\equiv\left\langle \left|\mathbf{w}(\mathbf{x}+\mathbf{r})-\mathbf{w}(\mathbf{x})\right|^{2}\right\rangle $
and likewise $F_{2}^{B}(r_{\parallel},r_{\perp})\equiv\left\langle \left|\mathbf{B}(\mathbf{x}+\mathbf{r})-\mathbf{B}(\mathbf{x})\right|^{2}\right\rangle $.
Here we adopt the procedure of \citet{ChoV2000}, but with some modifications.
Because our system is strongly inhomogeneous along the $y$ direction,
the structure functions increase rapidly as the displacement $\mathbf{r}$
moves along the $y$ direction. Therefore, we calculate the structure
functions for each $x-z$ plane, instead of using the full 3D space;
i.e. we only allow in-plane displacement with $\mathbf{r}\cdot\hat{\mathbf{y}}=0$.
Following \citet{ChoV2000}, the local magnetic field is defined as
the averaged field from the two points. However, to be consistent
with the allowed displacement, the parallel and perpendicular components
of the displacement are measured with respect to the in-plane component
of the local magnetic field. We compute the structure functions by
averaging over $10^{9}$ random pairs of points, whose $x$ coordinates
are within the current sheet region $-0.25\le x\le0.25$. The resulting
$F_{2}^{w}(r_{\parallel},r_{\perp})$ and $F_{2}^{B}(r_{\parallel},r_{\perp})$
are shown in Figure \ref{fig:Two-point-structure-functions} panels
(a) and (b) for $y=0$, and panels (d) and (e) for $y=0.002$, where
the contours reflect the shapes of eddies. These structure functions
clearly show turbulent eddies elongated along the local magnetic field
direction, with eddies at the $y=0.002$ plane more elongated than
eddies at the midplane $y=0$. These conclusions are qualitatively
consistent with earlier visual observation of Figure \ref{fig:Energy-fluctuations}.
{The contours of the structure function $F_{2}^{w}(r_{\parallel},r_{\perp})$
can also be used to infer the parameter $\chi\equiv k_{\perp}v_{l}/k_{\parallel}V_{A}\sim r_{\parallel}(F_{2}^{w})^{1/2}/r_{\perp}V_{A}$
as an indicator of the strength of nonlinear interaction. Here the
wave number parallel to the local magnetic field $k_{\parallel}$
is proportional to the semi-major axis $r_{\parallel}$ of a contour,
and likewise the perpendicular wave number $k_{\perp}$ is proportional
to the semi-minor axis $r_{\perp}$. Velocity fluctuation at the corresponding
length scales is given by $v_{l}\sim(F_{2}^{w})^{1/2}$, as the plasma
density $\rho\simeq1$. It can be inferred from Figure \ref{fig:Two-point-structure-functions}
(a) and (d) that $\chi$ is an $O(1)$ quantity over a broad range
of scales, indicating that the system is in a strongly nonlinear regime. }

{Structure function diagnostics also allow us to make
comparison with an important prediction of the Goldreich \& Sridhar
(GS) theory of incompressible MHD turbulence \citep{GoldreichS1995,GoldreichS1997},
namely that eddies become increasingly more anisotropic at smaller
scales. More precisely, GS theory predicts a scale-dependent anisotropy
relation $k_{\parallel}\sim k_{\perp}^{2/3}$, which is based on the
assumption of critical balance, i.e. the condition $k_{\parallel}V_{A}\sim k_{\perp}v_{l}$.
The scale-dependent anisotropy relation $k_{\parallel}\sim k_{\perp}^{2/3}$
has been confirmed by \citet{ChoV2000} by using two-point structure
function diagnostics. Here we repeat their procedure by plotting the
relationship between the semi-minor axis $r_{\perp}\sim1/k_{\perp}$
and the semi-major axis $r_{\parallel}\sim1/k_{\parallel}$ of contours
of structure functions. The results are shown in Figure \ref{fig:Two-point-structure-functions}
(c) and (d) for $y=0$ and $y=0.002$, respectively. The two dashed
lines in each panel represent the relations $k_{\parallel}\sim k_{\perp}$
(scale-independent) and $k_{\parallel}\sim k_{\perp}^{2/3}$ (GS theory),
for reference. In both panels, the relationships between $r_{\parallel}$
and $r_{\perp}$ appear to be more consistent with the scale-independent
relation $k_{\parallel}\sim k_{\perp}$ than the GS theory $k_{\parallel}\sim k_{\perp}^{2/3}$.
Therefore, we conclude that in this self-sustained turbulent system,
the eddy anisotropy is nearly scale-independent. }

\section{Discussion and Conclusion\label{sec:Discussion-and-Conclusion}}

In conclusion, our simulation results indicate that 3D plasmoid instabilities
in a reconnection layer can indeed lead to a self-sustained turbulent
state. This state is qualified as turbulent because it exhibits key
ingredients of turbulence, namely, energy cascade and development
of an extended inertial range. In addition, an important feature of
MHD turbulence, namely anisotropy of eddies with respect to local
magnetic field, is also observed. However, the turbulent state is
also highly inhomogeneous, therefore the applicability of conventional
MHD turbulence theories or phenomenologies becomes questionable. In
particular, we find the eddy anisotropy to be nearly scale-independent,
in contrast to the prediction of $k_{\parallel}\sim k_{\perp}^{2/3}$
by the Goldreich \& Sridhar theory. This discrepancy may be attributed
to several factors: (1) The background field is strongly sheared.
In our simulation the mean magnetic field rotates approximately 90
degrees across the reconnection layer, whereas most MHD turbulent
theories assume the presence of a strong uniform guide field or no
guide field at all. (2) Difference in the mechanism of energy cascade.
In the present case current sheet fragmentation caused by plasmoid
instabilities may play important roles in energy cascade, whereas
conventional MHD turbulence theories usually assume that energy cascade
is caused by interaction between counter propagating Alfv\'en waves.
(3) The turbulence is embedded in bi-directional Alfv\'enic mean
outflow jets, therefore disturbances in the reconnection layer will
be ejected in Alfv\'enic time scales. This distinct feature may interfere
with the energy cascade process and could be the reason why the inertial
range power-law spectra in our simulation are steeper than in most
homogeneous MHD turbulence situations. These considerations suggest
the necessity to develop new phenomenologies to account for this type
of turbulence that spontaneously arises in a reconnection layer.{
Recent theoretical work \citep{TerryAFFH2012} suggests that energy
spectrum in inhomogeneous turbulence driven by instabilities may not
be a pure power law, but a power law multiplied by an exponential
fall-off. We are currently investigating whether our numerical results
can be better explained by this new theoretical framework.}

In the present study, we find that once fully developed, 3D turbulent
reconnection rate is similar to 2D plasmoid-dominated reconnection
rate ($\sim0.01V_{A}B$), which is approximately an order of magnitude
slower than the fastest rate reported in an earlier study where turbulence
is driven by external forcing \citep{KowalLVO2009}. Even though the
scenarios of speeding up reconnection in 2D and 3D appear to be quite
different, the fact that they achieve approximately similar reconnection
rate is an encouraging finding and certainly needs to be further examined
in simulations with higher Lundquist numbers. At the present time
it has been relatively well-established that the reconnection rate
in 2D plasmoid-dominated reconnection is nearly independent of the
Lundquist number $S$. This conclusion is supported by heuristic arguments
and has been numerically tested for a wide range of Lundquist numbers
up to $S=10^{7}$ \citep{HuangB2010,UzdenskyLS2010,LoureiroSSU2012,HuangB2012,HuangB2013}.
Whether this conclusion remains valid in 3D is an important open question,
left to future work. A related important question is how fragmented
current sheet widths scale with resistivity $\eta$. In 2D the fragmented
current sheet widths follow a steep scaling $\delta\sim\eta$ such
that the electric field in secondary current sheets $\sim\eta B/\delta$
and becomes independent of $\eta$. This is in fact the main argument
why 2D plasmoid-dominated reconnection rate becomes independent of
$\eta$ \citep{HuangB2010,HuangB2013}. It becomes less clear whether
this feature will persist in the case of 3D turbulent reconnection
as the turbulent emf plays an important role in supporting the reconnecting
electric field. As has been discussed in the context of the reconnection
phase diagram by several authors \citep{HuangBS2011,JiD2011,CassakD2013},
the scaling of secondary current sheet widths may potentially impact
the criterion for transition from collisional to collisionless reconnection,
as transition typically takes place when the current sheet width becomes
smaller than kinetic scales such as ion skin depth or ion gyro radius. 

Even though the scaling of secondary current sheet widths and the
criterion for transition from collisional to collisionless reconnection
are uncertain at the present time, it is likely that in some astrophysical
applications dissipation may take place at kinetic scales as turbulence
cascades down to smaller scales. An important question is: will kinetic
physics at small scales feed back to large scales and alter the conclusion
from MHD simulations? Comparing our MHD results with those obtained
from collisionless particle-in-cell (PIC) simulation \citep{DaughtonRKYABB2011},
we note that oblique tearing modes play a similar role in developing
self-generated turbulent reconnection in both cases. A more recent
study \citep{DaughtonNKRL2014} also shows that 2D and 3D PIC simulations
yield similar reconnection rates. A noticeable difference between
the MHD and PIC simulations is the aspect ratio of turbulent region.
The turbulent region in PIC simulation is considerably broader, whereas
that in MHD simulation is more elongated. However, 3D PIC simulations
are limited to relative small system sizes typical of tens of ion
skin depths, therefore lack sufficient separation between large and
small scales. It remains an open question how the aspect ratio of
the turbulent region may change as collisionless PIC simulations scale
up to larger system sizes. 

In recent years, there have been  attempts to determine post-CME
(coronal mass ejection) current sheet thickness by using UVCS (UltraViolet
Coronagraph Spectrometer) and LASCO (Large Angle and Spectrometric
Coronagraph) observations \citep{CiaravellaR2008,LinLKR2009}. These
studies usually find the current sheet thickness to be significantly
broader than classical or anomalous resistivity would predict, and
turbulence has been suggested as a possible explanation. Using UVCS
observation for the 2003 November 4 CME event, \citet{CiaravellaR2008}
estimate the current sheet thickness to be within the range $0.04$
-- $\mbox{0.08}R_{\odot}$ at lower latitude ($\sim1.5R_{\odot}$),
whereas the current sheet length may be estimated from LASCO images
as approximately $4R_{\odot}$. That gives an estimate of the aspect
ratio to be within the range $50$ -- $100$. The study by \citet{LinLKR2009}
also finds similar current sheet thickness at lower latitude for other
events, while the thickness tends to increase at higher latitude.
Taking this into account, the estimated aspect ratio ($50$ -- $100$)
should be regarded as an upper bound. It should be noted that neither
UVCS nor LASCO measure the electric current density directly; instead,
they measure the temperature or density enhancement in the sheet-like
structure. If we assume the observed thickness is a consequence of
turbulence (e.g. due to turbulence heating and mixing), preliminary
comparison may be made with our simulation results. In the fully developed
turbulent state of our simulation, the thickness of the turbulent
region is approximately $0.02$ (Figure \ref{fig:EkEm1d}) while the
reconnection layer length is approximately $0.8$, therefore the aspect
ratio is approximately $40$, which is not inconsistent with the estimated
aspect ratio from UVCS observation. We should point out that our simulation
and above mentioned post-CME current sheets are of very different
global configurations and plasma parameter regimes, and important
effects such as plasma heating and thermal conduction are not included
in our model, therefore a direct comparison may not be possible. Future
work is needed to further assess whether turbulent broadening can
account for the observed thickness of post-CME sheet-like structure.

\acknowledgements{\textcolor{black}{We thank an anonymous referee for many insightful
suggestions. This work is supported by the National Science Foundation,
Grant Nos. PHY-0215581 (PFC: Center for Magnetic Self-Organization
in Laboratory and Astrophysical Plasmas), AGS-1331784, AGS-1338944,
and AGS-}1460169, and\textcolor{black}{{} NASA Grant Nos. NNX09AJ86G
and NNX10AC04G. Computations were performed with supercomputers at
the Oak Ridge Leadership Computing Facility and the National Energy
Research Scientific Computing Center. }}

\end{document}